\documentclass[aps,twocolumn,showpacs]{revtex4}
\usepackage{graphicx}
\usepackage{dcolumn}
\usepackage{bm,amsmath,amssymb}

\begin{document}

\title{Averaged null energy condition in Loop Quantum Cosmology}
\author{Li-Fang Li}
  \affiliation{Department of Physics, Beijing Normal University, Beijing 100875, China}
\author{Jian-Yang Zhu}
\thanks{Author to whom correspondence should be addressed}
  \email{zhujy@bnu.edu.cn}
  \affiliation{Department of Physics, Beijing Normal University, Beijing 100875, China}
\date{\today}

\begin{abstract}
Wormhole and time machine are very interesting objects in general
relativity. However, they need exotic matters which are impossible
in classical level to support them. But if we introduce the quantum
effects of gravity into the stress-energy tensor, these peculiar
objects can be constructed self-consistently. Fortunately, loop
quantum cosmology (LQC) has the potential to serve as a bridge
connecting the classical theory and quantum gravity. Therefore it
provides a simple way for the study of quantum effect in the
semiclassical case. As is well known, loop quantum cosmology is very
successful to deal with the behavior of early universe. In the early
stage, if taken the quantum effect into consideration, inflation is
natural because of the violation of every kinds of local energy
conditions. Similar to the inflationary universe, the violation of
the averaged null energy condition is the necessary condition for
the traversable wormholes. In this paper, we investigate the
averaged null energy condition in LQC in the framework of effective
Hamiltonian, and find out that LQC do violate the averaged null
energy condition in the massless scalar field coupled model.
\end{abstract}

\pacs{04.60.Pp,98.80.Qc,67.30.ef}
\maketitle

\section{Introduction}

The wormhole and the time machine are attractive objects in
general relativity and have been ever used as allurements to
attract young students to research general relativity
\cite{thorne88}. And they are always active research fields in
general relativity \cite{lobo07}. The deducing stress-energy
tensor components of these exotic spacetime through Einstein's
equation must violate all the known pointwise energy condition,
which is forbidden in classical general relativity. In contrast,
the energy condition violation can be easily met in quantum field
because of quantum fluctuation
\cite{klinkhammer91,epstein65,zeldovich71}. For example, the
Casimir vacuum for the electromagnetic field between two perfectly
conducting plates has a negative local energy density
\cite{casimir}; squeezed states of light can entail negative
energy densities \cite{braunstein}. But these quantum effects are
always confined to an extremely thin band \cite{roman04}. On the
other hand, the topological censorship theorem proved by Friedman,
Schleich, and Witt \cite{friedman} implies that, the existence of
macroscopic traversable wormholes requires the violation of the
averaged null energy condition (ANEC). ANEC can be stated as
\begin{eqnarray}
\int_\gamma T_{\mu\nu}k^\mu k^\nu dl\geq0, \label{energy}
\end{eqnarray}
where the integral is along any complete, achronal null geodesic
$\gamma $ , $k^\mu $ denotes the geodesic tangent, and $l$ is an
affine parameter.

If we take the effect of quantum gravity into consideration, the
quantum effect will enter stress-energy tensor. Then every kind of
energy condition violation is easily satisfied with respect to
effective stress-energy tensor which is reduced from metric
through Einstein's equation. Based on semi-classical gravitational
analysis, many self-consistent wormhole solutions have been found
\cite
{sushkov92,hochberg97,barcelo99,khusnutdinov03,garattini05,garattini07}.
As a quantum gravitational theory, loop quantum gravity
{(LQG)}\cite {ashtekar04,Smolin,rovelli98,thiemann01} is a
non-perturbative and background independent quantization of
gravity. The application of the techniques of {LQG} to the
cosmological sector is known as loop quantum cosmology {(LQC)}.
Some of the main features of LQG such as discreteness of spatial
geometry are inherited in {LQC}. {LQC} has the potential to serve
as a bridge connecting the classical theory and quantum gravity.
Therefore it provides a simple way for the study of quantum effect
in the semiclassical case. It exhibits an overwhelming strength in
solving the fundamental problems of the universe. A major success
of LQC is that it resolves the problem of classical singularities
both in an isotropic model \cite {bojowald01} and in a less
symmetric homogeneous model \cite{bojowald03}, a result that
depending crucially on the discreteness of the theory. It has been
shown that non-perturbative modification of the matter Hamiltonian
leads to a generic phase of inflation
\cite{bojowald02,date05,xiong1}. Furthermore, in this successful
quantum gravitational theory, another point worthy of being paid
some attention to is energy condition violation
\footnote{Physically speaking, the Einstein equations in LQC are
modified, while the stress-energy tensor $T_{\mu\nu}$ is
unchanged. But mathematically, we can move the terms modified by
LQC to the side of stress-energy tensor, and call them the quantum
effect of stress-energy tensor. This view point will bring some
advantages and make some previous analysis result still valid. For
an example, the proof of censorship theorem in \cite{friedman}
used the Einstein equations extensively. But it does not care
what's the form of matter. Instead, the geometric quantities are
very important to the proof. If we just put the modified terms
into the stress-energy tensor, the proof in that reference is
still valid. In this paper we take this view point.}. We do find
kinds of energy condition violation. For example, in loop quantum
cosmology non-perturbative modification to a scalar matter field
at short scales induces a violation of the strong energy condition
\cite{xiong2} and results in inflation
\cite{bojowald02,date05,xiong1}. Given so many local energy
condition violated by loop quantum cosmology and motivated by the
topological censorship theorem, we investigate that whether the
averaged null energy condition is violated also in LQC. The
difference between the ANEC and other energy conditions such as
strong energy condition mentioned above is that the ANEC is an
integral along any complete null-like geodesic, not be confined to
the neighborhood of a certain point of the space-time. For a
system without symmetry, it is a very complicated issue, making it
almost impossible to calculate. But in the context of LQC, taking
the cosmological principle into consideration, it will give us an
exact result. The cosmological principle states that the Universe
appears the same in every direction form every point in space.
This allows us to get a precise result and this result can provide
a hint for researching the wormholes in LQG and testing the
validity of LQG. In this paper, we assume that all the quantum
effect in LQC can be described approximately by effective loop
quantum cosmology. Through our calculation, we find that LQC do
violate the averaged null energy condition in the massless scalar
field coupled model.

This paper is organized as follows. First, we briefly review the
framework of effective loop quantum cosmology in Sec. \ref{Sec.2},
and introduce an exactly solvable model containing a massless scalar
field in Sec. \ref{Sec.3}. Then in Sec. \ref{Sec.4}, we investigate
the averaged null energy condition in this exactly solvable model.
At last in Sec. \ref{Sec.5}, we give some discussion and implication
of our results. In this paper we adopt $c=\hbar=G=1$.

\section{framework of effective LQC}

\label{Sec.2} In order to deduce the effective stress-energy tensor
conveniently, let us begin with the classical Hamiltonian description of the
universe. For simplicity, we only consider flat universe in this work. Under
the assumption of the cosmological principle, the metric of spacetime is
described by FRW metric
\[
ds^2=-dt^2+a^2\left( dx^2+dy^2+dz^2\right) ,
\]
where $a$ is the scale factor of the universe, only depending on $t$ due to
the homogeneity of our universe. The classical Hamiltonian for this system
is given by
\begin{equation}
H_{cl}=-\frac 3{8\pi \gamma ^2}\sqrt{p}c^2+H_M\left( p,\phi \right) .
\end{equation}
Here we have adopted the variables in the form of loop quantum
gravity. The phase space is spanned by coordinates $c=\gamma
\dot{a}$, being the gravitational gauge connection, and $p=a^2$,
being the densitized triad. Note that the Gaussian constraint in
loop formalism implies that changing $p$ to $-p$ will lead to the
same physical results, which can be seen obviously from the
relationship $p=a^2$. In this paper we fix this gauge freedom by
$p>0$. $\gamma $ is the Barbero-Immirzi parameter. $H_M$ denotes the Hamiltonian of matter part and $%
\phi $ means matter field. The dynamical equations together with
Hamiltonian constraint $H_{cl}=0$ which corresponds to Einstein's
equations reduce to the following Friedmann and Raychaudhuri
equations
\begin{eqnarray}
H^2 &=&\frac{8\pi }3\rho , \\
\frac{\ddot{a}}a &=&\dot{H}+H^2=-\frac{4\pi }3\left( \rho
+3P\right) .
\end{eqnarray}
Here the energy density $\rho $ and pressure $P$ of matter are defined \cite
{hossain05} as
\begin{eqnarray}
\rho =a^{-3}H_M ,\ P =-\frac 13a^{-2}\frac{\partial H_M }{
\partial a},\label{definition}
\end{eqnarray}
which is consistent with the form of stress-energy tensor for ideal
fluid
\begin{eqnarray}
T_{\mu \nu } &=&\rho U_\mu U_\nu +P\left( g_{\mu \nu }+U_\mu U_\nu \right)
\nonumber \\
&=&\rho \left( dt\right) _\mu \left( dt\right) _\nu   \nonumber \\
&&+a^2P\left[ \left( dx\right) _\mu \left( dx\right) _\nu +\left( dy\right)
_\mu \left( dy\right) _\nu +\left( dz\right) _\mu \left( dz\right) _\nu
\right] ,\nonumber \\
&&
\end{eqnarray}
where $U_\mu =(1,0,0,0)$ is the comoving observer of matter field which
corresponds to the comoving observer of the universe naturally.

Correspondingly, the effective Hamiltonian in LQC is given by \cite
{ashtekar06,mielczarek08}
\begin{equation}
H_{eff}=-\frac 3{8\pi \gamma ^2\bar{\mu}^2}\sqrt{p}\sin ^2\left( \bar{\mu}%
c\right) +H_M\left( p,\phi \right) .  \label{hamilton}
\end{equation}
The variable $\bar{\mu}$ corresponds to the dimensionless length of the edge
of the elementary loop and is given by
\begin{equation}
\bar{\mu}=\xi p^{-1/2},
\end{equation}
where $\xi $ is a constant $\xi >0$ and depends on the particular scheme in
the holonomy corrections. In this paper we take $\bar{\mu}$-scheme, which
gives
\begin{equation}
\xi ^2=2\sqrt{3}\pi \gamma l_p^2,
\end{equation}
where $l_p$ is Planck length.

Similarly, the canonical equations and the Hamiltonian constraint
give out modified Friedmann and Raychaudhuri equations
\begin{eqnarray}
H^2 &=&\frac{8\pi }3\rho \left( 1-\frac \rho {\rho _c}\right) , \label{m_Friedmann}\\
\dot{H}+H^2 &=&-\frac{4\pi }3\left[ \rho \left( 1-\frac \rho {\rho _c}%
\right) +3P\left( 1-\frac{2\rho }{\rho _c}\right) -3\frac{\rho ^2}{\rho _c}%
\right] ,\nonumber \\
&&\label{m_Raychaudhuri}
\end{eqnarray}
where $\rho _c=\frac 3{8\pi \gamma ^2\xi ^2}$ is the critical energy density
coming from the quantum effect in loop quantum cosmology. These modified
equations give us the effective energy density and the effective pressure
reduced by Einstein's equation as
\begin{eqnarray}
\rho _{eff} &=&\rho \left( 1-\frac \rho {\rho _c}\right) , \\
P_{eff} &=&P\left( 1-\frac{2\rho }{\rho _c}\right) -\frac{\rho ^2}{\rho _c}.
\end{eqnarray}
The introduction of these effective quantities makes the reduced
stree-energy tensor in the form of the ideal fluid again.

\section{An exactly solvable model}

\label{Sec.3} Now following closely \cite{mielczarek08}, we
consider a universe containing a massless scalar field. Then the
Hamiltonian for the matter part in equation (\ref{hamilton}) can
be written as
\[
H_M(p,\phi )=\frac 12\frac{p_\phi ^2}{p^{3/2}},
\]
where $p_\phi $ is the conjugate momentum for the scalar field
$\phi $. The complete equations of motion for the universe
containing a massless scalar field are
\begin{equation}
\left\{
\begin{array}{c}
\dot{c}=-\frac 1\gamma \frac \partial {\partial p}\left(
\sqrt{p}\left[
\frac{\sin \left( \bar{\mu}c\right) }{\bar{\mu}}\right] ^2\right) -\frac{%
\kappa \gamma }4\frac{p_\phi ^2}{p^{5/2}}, \\
\dot{p}=\frac 2\gamma
\frac{\sqrt{p}}\mu \sin \left( \bar{\mu}c\right) \cos \left(
\bar{\mu}c\right) ,
\end{array}
\right.   \label{p-c}
\end{equation}
and
\begin{equation}
\left\{
\begin{array}{c}
\dot{\phi}=p^{-\frac 32}p_\phi , \\
\dot{p}_\phi =0,
\end{array}
\right.   \label{phi-pphi}
\end{equation}
where $\kappa =8\pi $. In addition, the Hamiltonian constraint $H_{eff}=0$
becomes
\begin{equation}
\frac 3{8\pi \gamma ^2\bar{\mu}^2}\sqrt{p}\sin ^2\left( \bar{\mu}c\right) =%
\frac 12\frac{p_\phi ^2}{p^{3/2}}.  \label{eq_h}
\end{equation}
Combining equations (\ref{p-c}) and (\ref{eq_h}) we obtain
\begin{equation}
\left( \frac{dp}{dt}\right) ^2=\Omega _1p^{-1}-\Omega _3p^{-4},
\label{equation}
\end{equation}
with $\Omega _1=\frac 23\kappa p_\phi ^2$ and $\Omega _3=\frac 19\kappa
^2\gamma ^2\xi ^2p_\phi ^4$. Note that from equation (\ref{phi-pphi}), $%
p_\phi $ is a constant which characterizes the scalar field in the system.
To solve equation (\ref{equation}) we introduce a new dependent variable $u$
in the form
\begin{equation}
u=p^3.
\end{equation}
With use of the variable $u$ the equation (\ref{equation}) becomes
\begin{equation}
\left( \frac{du}{dt}\right) ^2=9\Omega _1u-9\Omega _3,
\end{equation}
and has a solution
\begin{equation}
u=\frac{\Omega _3}{\Omega _1}+\frac 94\Omega _1t^2-\frac 92\Omega _1C_1t+%
\frac 94\Omega _1C_1^2,
\end{equation}
where $C_1$ is a constant of integration. We can choose $C_1=0$ through
coordinate freedom. Then the solution for $p$ is
\begin{equation}
p=\left[ \frac{\Omega _3}{\Omega _1}+\frac 94\Omega _1t^2\right] ^{1/3}.
\label{solution}
\end{equation}

\section{the averaged null energy condition in LQC}

\label{Sec.4}Based on the above discussion, we come to calculate the
averaged null energy condition in the context of LQC. Because of the
homogeneity of the universe described by the FRW metric, the null
geodesic curves pass through different spacial points are the same.
So, to investigate the ANEC, we need only consider one of the null
geodesic lines pass any point in space. Due to the isotropy of FRW
metric, the null geodesic curves passing through the same point in
different directions are also the same. Therefore, our problem
reduces to test along any null geodesic line. Specifically, we
consider this null geodesic line generated by vectors
\[
\left( \frac \partial {\partial t}\right) ^\mu +\frac 1a\left( \frac \partial
{\partial x}\right) ^\mu .
\]
According to definition $(\frac \partial {\partial l})^\mu \nabla _\mu (%
\frac \partial {\partial l})^\nu =0$ we can reparameterize it with affine
parameter $l$ to get
\[
k^\mu =\left( \frac \partial {\partial l}\right) ^\mu =\frac 1a\left( \frac %
\partial {\partial t}\right) ^\mu +\frac 1{a^2}\left( \frac \partial {%
\partial x}\right) ^\mu .
\]
Then we can get the relation between $t$ and affine parameter $l$
\[
t=\frac la.
\]

For the considered universe containing a massless scalar field, the energy
density and the pressure of matter can be expressed as
\begin{equation}
\rho =\frac 12\dot{\phi}^2,
\end{equation}
\begin{equation}
P=\frac 12\dot{\phi}^2,
\end{equation}
according to the definition (\ref{definition}). In the context of LQC,
taking the quantum effects into consideration, the energy density and the
pressure of matter reduce to the effective forms
\begin{equation}
\rho _{eff}=\frac 12\dot{\phi}^2\left( 1-\frac 12\frac{\dot{\phi}^2}{\rho _c}%
\right) ,
\end{equation}
\begin{equation}
P_{eff}=\frac 12\dot{\phi}^2\left( 1-\frac{\dot{\phi}^2}{\rho _c}\right) -%
\frac 14\frac{\dot{\phi}^4}{\rho _c}.
\end{equation}
Since the stress-energy tensor reduced from the Einstein's equations takes
ideal fluid form in terms of the effective quantities
\begin{eqnarray}
T_{\mu \nu }^{eff} &=&\rho _{eff}\left( dt\right) _\mu \left(
dt\right) _\nu
+a^2P_{eff}  \nonumber \\
&&\times \left[ \left( dx\right) _\mu \left( dx\right) _\nu +\left(
dy\right) _\mu \left( dy\right) _\nu +\left( dz\right) _\mu \left(
dz\right) _\nu \right] \nonumber \\
&&
\end{eqnarray}
the average null energy condition (\ref{energy}) becomes
\begin{eqnarray}
\int_\gamma T_{\mu \nu }^{eff}k^\mu k^\nu dl &=&\int_{-\infty }^\infty \frac %
1a\left( \rho _{eff}+P_{eff}\right) dt  \nonumber \\
&=&\int_{-\infty }^\infty \frac 1a\left( \dot{\phi}^2-\frac{\dot{\phi}^4}{%
\rho _c}\right) dt  \nonumber \\
&=&p_\phi ^2\int_{-\infty }^\infty (p^{-7/2}-p^{-13/2}\frac{p_\phi ^2}{\rho
_c})dt.  \nonumber
\end{eqnarray}
In the last line we have used equation (\ref{phi-pphi}) and the relationship
between $p$ and $a$. Substitute the exact solution (\ref{solution}) in above
equation, we get
\begin{equation}
\int_\gamma T_{\mu \nu }^{eff}k^\mu k^\nu dl=-\frac{\Gamma (5/6)\Gamma (2/3)%
}{7\rho _c\Omega _I(\frac{\Omega _{III}}{\Omega _I})^{13/6}\sqrt{\frac \pi {%
\Omega _{III}}}}p_\phi ^4,
\end{equation}
where $\Gamma $ is the gamma function. From above result it is obvious that
\begin{equation}
\int_\gamma T_{\mu \nu }^{eff}k^\mu k^\nu dl<0.
\end{equation}
From the above discussion, we can clearly see that, in addition to violation
of every kind of local energy condition, loop quantum cosmology also
violates the averaged null energy condition. This averaged null energy
condition violation results from quantum effect completely.

\section{summary and discussion}

\label{Sec.5} Given many kinds of local energy condition violation
in loop quantum cosmology and motivated by the topological
censorship theorem which rules out traversable wormholes in
spacetime where the averaged null energy condition is satisfied,
we investigate this kind of nonlocal energy condition in the
context of LQC. Our analyses are based on a flat universe
containing a massless scalar field. This model can be solved
analytically. With the help of the analytical solution and taking
advantage of the homogenous and isotropic symmetry of universe, we
calculate the average of energy directly. Our result is
interesting. Although the quantum correction is focused on the
early universe around Planck scale, the correction is so strong
that it makes the universe violate the null averaged energy
condition. Mathematically we have written the modified Einstein
equations in LQC in original Einstein equations form (refer to
Eqs.~(\ref{m_Friedmann})-(\ref{m_Raychaudhuri})) but with
effective stress-energy tensor. This form of equations makes the
proof of the censorship theorem in \cite{friedman} valid also. So
the ANEC (for the original stress-energy tensor instead of the
effective one) argument can not forbid existence of wormhole once
the Loop Quantum Gravity effects are taken into account. But we do
not expect the existence of wormhole in LQC, because it is too
symmetric to support wormholes \footnote{We thank our referee for
pointing out this to us.}. On the other hand, LQC adopts the
essence of LQG, so our result can shed some light on the ANEC of
LQG. And we hope this result can give some hints for looking for
wormhole solutions in the LQG theory. These interesting objects
will provide some gedanken-experiments to test our quantum gravity
theory.

\acknowledgements

The work was supported by the National Natural Science Foundation of
of China (No.10875012).


\begin{thebibliography}{99}
\bibitem{thorne88}M. Morris and K. Thorn, Am. J. Phys. {\bf 56}, 395
(1988).

\bibitem{lobo07}F. Lobo, arXiv:0710.4474.

\bibitem{klinkhammer91}G. Klinkhammer, Phys. Rev. D {\bf 43}, 2542
(1991).

\bibitem{epstein65}H. Epstein, V. Glaser, and A. Jaffe, Nuovo Cimento
{\bf 36}, 1016 (1965).

\bibitem{zeldovich71}Y. Zeldovich and L. Pitaevsky, Commun. Math.
Phys. {\bf 23}, 185 (1971).

\bibitem{casimir}H. Casimir, Proc. K. Ned. Akad. Wet. B {\bf 51},
793 (1948).

\bibitem{braunstein}Unpublished work by S. Braunstein, described in \cite
{thorne88}.

\bibitem{roman04}T. A. Roman, arXiv:gr-qc/0409090.

\bibitem{friedman}J. Friedman, K. Schleich, and D. M. Witt, Phys. Rev.
Lett. {\bf 71}, 1486 (1993).

\bibitem{sushkov92}S. V. Sushkov, Phys. Lett. A {\bf 164}, 33 (1992).

\bibitem{hochberg97}D. Hochberg, A. Popov and S. V. Sushkov, Phys. Rev.
Lett. {\bf 78}, 2050 (1997).

\bibitem{barcelo99}C. Barcelo and M. Visser, Phys. Lett. B {\bf 466}
127 (1999).

\bibitem{khusnutdinov03}N. R. Khusnutdinov, Phys. Rev. D 67, 124020 (2003).

\bibitem{garattini05}R. Garattini, Class. Quant. Grav. {\bf 22} 1105
(2005).

\bibitem{garattini07}R. Garattini and F. Lobo, Class. Quant. Grav.
{\bf 24}, 2401 (2007).

\bibitem{ashtekar04}A. Ashtekar and J. Lewandowski, Class. Quantum Grav.
21, R53 (2004).

\bibitem{Smolin}L. Smolin, An invitation to loop quantum gravity,
hep-th/0408048.

\bibitem{rovelli98}C. Rovelli, Liv. Rev. Rel. 1, 1 (1998); {\it Quantum
Gravity}, (Cambridge Univ. Press, Cambridge, 2004).

\bibitem{thiemann01}T. Thiemann, {\it Modern canonical quantum general
relativity}, (Cambridge Univ. Press, Cambridge, 2004).

\bibitem{bojowald01}M. Bojowald, Phys. Rev. Lett. {\bf 86}, 5227 (2001).

\bibitem{bojowald03}M. Bojowald, Class. Quantum Grav. {\bf 20}, 2595
(2003).

\bibitem{bojowald04a}M. Bojowald, G. Date, Phys. Rev. Lett. {\bf 92}, 071302 (2004).

\bibitem{bojowald04b}M. Bojowald, Class. Quantum Grav. {\bf 21}, 3541
(2004).

\bibitem{bojowald02}M. Bojowald, Phys. Rev. Lett. {\bf 89}, 261301
(2002).

\bibitem{date05}G. Date and G. M. Hossain, Phys. Rev. Lett. {\bf 94},
011301 (2005).

\bibitem{xiong1}Hua-Hui Xiong and Jian-Yang Zhu, Phys. Rev. D {\bf 75%
}, 084023 (2007).

\bibitem{xiong2}Hua-Hui Xiong, Jian-Yang Zhu, Inter. J. Mod. Phys. A
, {\bf 22}, 3137 (2007).

\bibitem{hossain05}G. Hossain, Classical Quantum Gravity 22, 2653 (2005).

\bibitem{ashtekar06}A. Ashtekar, T. Pawlowski, and P. Singh, Phys.
Rev. D {\bf 73}, 124038 (2006); {\bf 74}, 084003 (2006).

\bibitem{mielczarek08}J. Mielczarek, T. Stachowiak, M. Szydlowski,
Phys. Rev. D {\bf 77}, 123506 (2008); arXiv:: 0801.0502.
\end{thebibliography}
\end{document}